\documentclass[natbib]{svjour3}                     
\smartqed  
\usepackage{graphicx}
\usepackage{mathptmx}      
\newcommand{\kms}{km\,s$^{-1}$}
\newcommand{\ergs}{erg\,s$^{-1}$}
\newcommand{\arcsec}{{''}}
\newcommand{\pyr}{${\rm yr}^{-1}$}
\newcommand{\Einstein}{\emph{Einstein}}
\newcommand{\ROSAT}{\emph{ROSAT}}
\newcommand{\ASCA}{\emph{ASCA}}
\newcommand{\Chandra}{\emph{Chandra}}
\newcommand{\XMM}{\emph{XMM-Newton}}
\journalname{Space Science Reviews}
\newcommand{\thisjour}{{Space Sci.\ Rev.}}
\newcommand{\aipc}{{AIP Conf.\ Proc.}}
\newcommand{\aj}{{Astron.\ J.}}
\newcommand{\ana}{{A\&A}}
\newcommand{\apjss}{{Astrophys.\ J.\ Sup.\ Ser.}}
\newcommand{\apj}{{Astrophys.\ J.}}
\newcommand{\apjl}{{Astrophys.\ J.\ Let.}}
\newcommand{\apss}{{Astrophysics \& Space Science}}
\newcommand{\araa}{{Annu.\ Rev.\ Astro.\ Astrophys.}}
\newcommand{\aspc}{{ASP Conf.\ Series}}
\newcommand{\baas}{{Bull.\ Am.\ Astron.\ Soc.}}
\newcommand{\cqg}{{Class.\ Quantum Grav.}}

\newcommand{\lnp}{{Lecture Notes in Physics}}
\newcommand{\mnras}{{Mon.\ Not.\ R.\ Astron.\ Soc.}}
\newcommand{\pasj}{{Publ.\ Astron.\ Soc.\ Japan}}
\newcommand{\pasp}{{Publ.\ Astron.\ Soc.\ of the Pacific}}
\newcommand{\prl}{{Phys.\ Rev.\ Let.}}
\newcommand{\reprph}{{Rep.\ Prog.\ Phys.}}
\newcommand{\science}{{Science}}
\newcommand{\snronine}{\textit{SNR \& PWN in the Chandra Era}}
\newcommand{\spie}{{Proc.\ of SPIE}}
\begin{document}

\title{Kinematics of Supernova Remnants: Status of X-Ray Observations
}

\titlerunning{Kinematics of SNRs}        

\author{Daniel Dewey}

\authorrunning{Dewey} 

\institute{D. Dewey \at
              MIT Kavli Institute, Cambridge, MA 02139, USA \\
              \email{dd@space.mit.edu}           
}

\date{Received: date / Accepted: date}

\maketitle

\begin{abstract}
A supernova (SN) explosion drives stellar debris into the circumstellar material (CSM)
filling a region on a scale of parsecs with X-ray emitting plasma.
The velocities involved in supernova remnants (SNRs), thousands of \kms,
can be directly measured with medium and high-resolution X-ray spectrometers and add an
important dimension to our understanding of the last stages of the progenitor,
the explosion mechanism, and the physics of strong shocks.
After touching on the ingredients of SNR kinematics,
I present a summary of the still-growing measurement results from
SNR X-ray observations.  Given the advances in 2D/3D hydrodynamics, data
analysis techniques, and especially X-ray instrumentation, it is clear that
our view of SNRs will continue to deepen in the decades ahead.

\end{abstract}


\section{Introduction}
\label{intro}

In an instant, the relative stability and compact extent ($\sim 10^9$\,cm) of a white dwarf
or the core of a massive star are dramatically transformed,
by a thermonuclear or core-collapse supernova (SN) explosion, respectively. 
A cosmic moment later (hundreds of years), the expanding SN
debris, the ``ejecta'', has interacted with the pre-explosion circumstellar
medium (CSM)
and fills a region over a scale of order parsecs (1~pc = 3.086$\times10^{18}$~cm).
The ``outputs'' of the SN explosion may include a neutron star (NS) or
black hole (BH), as well as newly synthesized elements and their dispersal into the
interstellar medium (ISM).
The velocities involved in these supernova remnants (SNRs) are several
thousands of \kms, making them inherently dynamic objects.
Hence, measuring the spatial-velocity properties of SNRs adds an
important  third dimension to our view of them.
Because the bulk of the SNR material, closely connected to the
global hydrodynamics, is generally heated well above a
million degrees there is considerable emission in the X-ray band and
X-ray data play a unique role in understanding the SNR.

Of course, there is a tremendous amount of
information on SNRs and their kinetics provided by non-X-ray means.
The kinetic energy of the SN explosion is turned into radiation across the
electromagnetic (E/M) band: radio, sub-mm, IR, NIR, optical, UV, EUV, X-ray, $\gamma$-ray,
and on up to over 1~TeV; for example, see the figures in \citet{Araya10}.
Emission mechanisms include synchrotron, thermal lines and bremsstralung, and
high-energy non-thermal components \citep{Reynolds08b}.
Non-linear shock acceleration also leads to 
particle acceleration at SNRs and generation of cosmic rays
\citep[CRs,][]{Vink10}.
Multi-wavelength observations bring out the full complexity of SNRs.  As an early example, 
the SNR Doppler velocities from optical measurements
were much slower than the shock speeds needed to create the (then) newly
observed thermal X-rays, this led to the
suggestion of density variations of 5--10 times within the CSM \citep{Bychkov75}.
This was the beginning of an on going difference of kinematics seen in various
wavelengths and among different components of an SNR.
The instrumentation and techniques in non-X-ray bands are often ahead of
current X-ray capabilities, for example, optical H$\alpha$ measurements
of Balmer-dominated shocks directly
measure the post-shock proton thermal velocity.

Besides E/M radiation and emission of high-energy particles,
the SNe process has been observed through neutrinos
\citep[SN~1987A,][]{Hirata87}
and in future we expect their gravitational waves (GWs) to be detected,
e.g., with Advanced LIGO \citep{Ott09,Yakunin10}.
The study of SNe/SNRs also ties into other areas of
astrophysics including: the late stages of massive stars and nucleosynthesis \citep{Thielemann10},
the formation of black holes and neutron stars, pulsar wind nebulae
\citep[PWN,][]{Slane08}, 
cosmology through the use of thermonuclear SNe (``Type~Ia's'') as standard candles,
gamma-ray burst sources (GRBs) as a sub-class of
Type Ibc SNe \citep{Woosley06}, and GRB remnants \citep{Ramirez-Ruiz10}.

The aims of this article are to touch on the key
ingredients in SNR kinematics (\S\ref{sec:theory}), including the collisionless shock
and basic SNR structure, and then to summarize the direct X-ray kinematic measurements that
have been made to-date on SNe/SNRs (\S\ref{sec:obs}). I conclude
with a brief look at future X-ray instrumentation\
capabilities in the context of SNRs (\S\ref{sec:future}).

\section{Ingredients of SNR Kinematics}
\label{sec:theory}

\subsection{The Collisionless Shock}
\label{sec:shock}

A key ingredient in SNR dynamics is the strong (high Mach number) shock
which is ``collisionless'' in that the 
effect of the shock is carried out through E/M fields generated collectively
by the plasma rather than through discrete particle--particle collisions
\citep{Ghavamian07}.
A further characterization of the shock system is given by
the synonymous terms ``adiabatic'' and ``non-radiative'' to indicate
that no significant energy leaves the system.  In contrast, 
a ``radiative'' shock describes the case where significant, catastrophic cooling takes
place through E/M emission (generally into the optical), and
a ``cosmic-ray (CR) modified'' shock \citep{Decourchelle00,Vink10} indicates
that significant energy is leaving the system through
highly accelerated particles.

The main results from the kinematic point of view are the expressions for the post-shock
bulk velocity and temperature in terms of the shock velocity, the speed of the shock front
as it moves upstream into the unshocked, stationary medium (e.g., the CSM).
Assuming a usual $\gamma=5/3$ equation of state, the shocked
material picks up a velocity: 
\begin{equation}
v_{\rm bulk} ~=~ \big(1 - {{1}\over{\chi}}\big) ~ v_s~~~=~
{{3}\over{4}} ~ v_s 
\end{equation}
\noindent where $\chi = \rho_{\rm post-shock}/\rho_{\rm CSM}=4$
is the shock compression ratio.
The temperature of particles of mass $m_i$
in the shocked material just behind (or ``downstream'' of) the shock
is given by:
\begin{equation}
kT_i = (3/16)\,m_i v_s^2 \,,
\end{equation}
\noindent implying a very low post-shock electron-to-proton temperature
ratio, $T_e/T_p = 1/1836$.  However, there is strong
evidence that additional electron heating occurs at collisionless
shocks with $T_e/T_p$ ranging from 0.01 to 1.0 as $v_s$
decreases from 4000 to 400\,\kms\ \citep{Ghavamian07}.

In time, particle (``Coulomb'') collisions in the post-shock plasma will
bring the temperature of all species, including the free electrons,
to an equilibrium value:
\begin{equation}
\bar{kT}={{3}\over{16}}\mu v_s^2
\label{eq:kTshock}
\end{equation}
\noindent where $\mu$ is the mean mass per particle in the plasma.
Likewise, post-shock ions will be further ionized through electron collisions
producing a non-equilibrium ionization (NEI) evolution in the plasma
state \citep{Hamilton83,Borkowski01,Smith10}.
The NEI state may be approximately and conveniently expressed in terms
of the temperature and the ionization age, $\tau\equiv\int n_e dt$, at locations within the
plasma; these values can be used to calculate the expected X-ray emission
\citep{Borkowski01,Dwarkadas10}.  

From a kinetic observable point of view, we are interested in the
rms thermal motion along one dimension (e.g., the
line-of-sight) which, immediately post-shock, is given by:
\begin{equation}
\sigma_{\rm ps,therm}=\sqrt{{{kT_i}\over{m_i}}}={{\sqrt{3}}\over{4}}v_s 
\end{equation}
\noindent independent of $m_i$.
Once temperature equilibration has been established among the plasma
species, different ions will then have different thermal velocity dispersions:
\begin{equation}
\sigma_{i,{\rm therm}}=\sqrt{{{\bar{kT}}\over{m_i}}} =
\sqrt{\beta_{\rm CR}}~\sqrt{{{\mu}\over{m_i}}}{{\sqrt{3}}\over{4}}v_s 
\label{eq:sigtherm}
\end{equation}
\noindent and the thermal broadening then depends
on the mass of the ion emitting the line.  Note that the factor
$\sqrt{\mu/m_i}$ is generally less than 0.25 for N and heavier elements
in solar-abundance plasmas. 
The efficient production of cosmic rays at the shock may produce an even
greater effect on the gas heating:
``to the point where thermal X-ray emission is no longer expected'' \citep{Drury09}.
The value of $\bar{kT}$ is then reduced from the unmodified value
given in eq.\,(\ref{eq:kTshock}) by a factor $\beta_{\rm CR}$ \citep{Vink10},
which has been included in the equation above.

This thermal Doppler broadening can be compared with the broadening that
the bulk velocity introduces.  Assuming emission from
a complete thin spherical shell, we get a lineshape that is
uniform from $-v_{\rm bulk}$ to $+v_{\rm bulk}$; with a high-resolution
spectrometer this shape may be resolved
and the ``blue-velocity at zero intensity'' (BVZI) measures $v_{\rm bulk}$
directly \citep{Chugai94}.
At lower resolution it is
useful to express this as an equivalent line broadening.
Using the result that a
uniform distribution from $-a$ to $+a$ has an rms (i.e. one sigma) value of $a/\sqrt{3}$, we
get the effective bulk velocity broadening from the whole shell as
equivalent to a Gaussian with:
\begin{equation}
\sigma_{\rm shell}^{\rm equiv.} = {{v_{\rm bulk}}\over{\sqrt{3}}} =
{{\sqrt{3}}\over{4}}v_s
~~,  ~~~{\rm or}:~~
{\rm FWHM}_{\rm shell}^{\rm equiv.} = 2\sqrt{\ln(4)} ~ {{\sqrt{3}}\over{4}}v_s
\approx 1.02 ~ v_s ~.
\label{eq:sigshell}
\end{equation}
\noindent (Note that under these assumptions,
we can say that the FWHM is equal to the shock
velocity to reasonable accuracy.)  Compared with the thermal broadening, eq.\,(\ref{eq:sigtherm}),
we see that the bulk broadening will generally dominate the global linewidth.
Of course, if the analysis is spatially resolved then the bulk term can be
reduced by using data near the SNR rim. Likewise, for data restricted to the center
of the SNR the bulk contribution is enhanced, with
$\sigma_{\rm shell}^{\rm equiv.}\approx v_{\rm bulk}$.  These effects
seem to be seen in Tycho observations, \S\ref{sec:Ia} below.

\begin{figure*}[t]
\center{\includegraphics[width=1.0\textwidth]{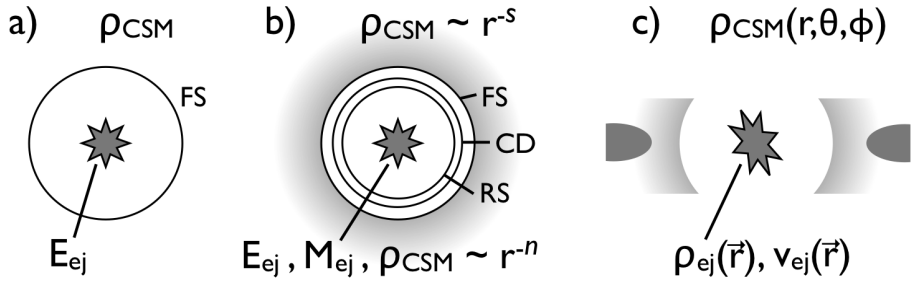}}
\caption{A physicist's progression of SNR kinematic models. a) The Sedov
solution depends only on the explosion energy and the ambient density and
is most appropriate at later stages of SNR evolution.
b) More realistic at early times, a self-similar solution
allows powerlaw radial profiles for both the
ejecta and CSM; this gives rise to the usual SNR structure with a
forward shock (FS), contact discontinuity (CD), and reverse shock (RS).
c) In a general case, at $t\approx 0$ the ejecta and/or CSM can have complex spatial and
velocity configurations and 2D/3D hydrodynamic simulation is required to determine the 
SNR formation and evolution.
\label{fig:geoms} }
\end{figure*}

\subsection{SNR Structure}
\label{sec:structure}

As a start, we can consider an SNR to be
an idealized explosion which sends an expanding shock front
into a medium, Figure~\ref{fig:geoms}a.
\citet{Sedov59} gives the radius of an ``intense explosion'', e.g, 
an atomic bomb explosion in New Mexico, as a function of time as:
\begin{equation}
r(t) ~\sim~ 1.15~ \big({{E_{\rm ej}}\over{\rho_{\rm CSM}}}\big)^{1/5} ~ t^{2/5} ~.
\end{equation}
\noindent Here the constant of proportionality is given 
for a strong, adiabatic shock into a medium that has
typical cosmic elemental abundances.
Setting $E_{\rm ej}$ to 10$^{51}$ \ergs, $\rho_{\rm CSM}=$1\,amu\,cm$^{-3}$
(1\,amu $\sim$\,1.66$\times10^{-24}$\,g), and $t=$\,1000\,yr,
gives $r\approx$\,8.5\,pc or an angular diameter
of 70\,\arcsec\ at the LMC (50~kpc).
For further application of the Sedov solution
to SNRs and their X-ray emission see \citet{Hamilton83} and 
\citet{Borkowski01}.

Implicit in the simple Sedov solution is the assumption that the mass of the
shocked material is only due to (or is at least dominated by) the mass
of the ``swept up'' ambient medium.  In the case of SNRs,
one can consider the Sedov solution as the {\it late-time limit}
when the swept-up CSM mass exceeds the SN ejecta mass and the SNR evolution
retains only vestiges of the initial ejecta mass and its distribution. 
Earlier in the ejecta-CSM interaction,
the characteristics of the ejected material must be
considered.

Self-similar solutions to describe this initial stage of
SNR interaction were independently derived by \citet{Chevalier82} and
\citet{Nadyozhin85}.
At this next level of fidelity, Figure~\ref{fig:geoms}b,
the density of the material ejected in the SN explosion 
can be described as a decreasing power-law with radius.
This ejecta behavior for core-collapse SN
\footnote{The power-law profile is useful for modeling Type~Ia's, however
an exponential profile may be more realistic for these \citep{Dwarkadas98,Dwarkadas00}.}
has been corroborated by analytic \citep{ChevalierSoker89} and
numerical calculations; realistic values of the power-law index
are expected between about 9 and 11 \citep{Matzner99}.
Because the expansion is homologous, $v(r)=r/t$,
the density can also be expressed as a power-law with velocity: $\rho \propto v^{-n}$. 
Since a power-law cannot be continued all the
way to the origin, as it would result in infinite mass and energy, the
density below a certain velocity is simply assumed to be constant
\footnote{Another possibility is to transition to a more shallow power-law
at small radii \citep{Chevalier94}; the constant-density case here is then
a subset of this with the inner index, $\delta$, set to 0.}. With this
assumption, the requirement of finite mass and energy then gives the constraint
$n>5$.  The parameters of the ejecta profile are directly related to the
ejecta mass and explosion energy, see eq.\,(2.1) of \citet{ChevalierLiang89}.

The interaction of this power-law ejecta density
profile with an ambient medium whose density is also a power-law with
radius, $\rho_{\rm CSM} \propto r^{-s}$\,,
leads to the formation of the standard double-shocked structure, Figure~\ref{fig:geoms}b.
The outer, forward shock (FS)
advances into the ambient medium, which could be a constant-density interstellar
medium ($s=0$) as expected for Type~Ia SNe, or a steady wind from the progenitor
star ($s=2$) in the case of a core-collapse SN. 
We assume that $v_{\rm CSM} \ll v_{\rm FS}$ so that the shock velocity is
the same as the observed FS motion: $v_{s,{\rm FS}}\approx v_{\rm FS}$.
A reverse shock (RS) is formed that
moves back into the ejecta in a Lagrangian sense. At the RS, the upstream
(unshocked) ejecta has been freely expanding and the observed velocity of the \emph{location}
of the RS, $v_{\rm RS}$, is not the shock velocity, $v_s$, of
\S\ref{sec:shock};
instead we have: $v_{s,{\rm RS}} = r_{\rm RS}/t \,-\,v_{\rm RS}$.

A contact discontinuity (CD) separates the shocked ejecta from the shocked
surrounding medium.  The entire shocked structure expands
outwards, with the ratios of the radii of the FS and RS
to the radius of the CD remaining constant.
The self-similar solution for the radius of the contact discontinuity has been
shown to increase as $t^{(n-3)/(n-s)}$ \citep{Chevalier82,Chevalier94}.
Note that for $n>5$ and $s\le 2$, the
exponent will always be larger than in the generalized Sedov
case, $2/(5-s)$.

This analytic
framework has assumed an adiabatic evolution, i.e., with no additional
energy sinks beyond the expansion.  However, it is able to accomodate
additional energy terms, as examples:
i) \citet{Nymark06} consider the case of a radiative RS, where the
radiative cooling is relevant and a cool dense shell forms at the CD,
and ii) \citet{Decourchelle00} show the effect on the kinematics of
including cosmic-ray generation, which changes the relative locations of the
RS, CD, and FS. 

The self-similar solution eventually breaks down, for example when the
RS reaches the plateau of the ejecta distribution.
The solution then deviates from the self-similar one and the
expansion parameter changes, decreasing towards the limiting Sedov value.
In practice this happens only when
the FS has swept-up a mass equal to several times the mass
of the ejected material \citep{Dwarkadas98}.

\subsection{Rayleigh-Taylor \& Kelvin-Helmholtz Instabilities}

The idealized 1D symmetric picture of the previous section
suffers from several further complications: the decelerating contact discontinuity
is always Rayleigh-Taylor (R-T) unstable. 
The contact discontinuity is decelerated by the continual sweeping up
of the ambient medium by the FS. The deceleration of the
shocked ejecta by the shocked ambient medium leads to conditions
favoring the rise of the R-T instability at the
contact discontinuity. R-T fingers of ejecta can be seen
penetrating the shocked ambient medium in numerical simulations \citep{Chevalier92,Dwarkadas00}.
The tips of these fingers usually show
bulbous structures due to shear flow around the heads, leading to the
onset and growth of Kelvin-Helmholtz instabilities. This can lead to
mixing of heavy-metal ejecta into the shocked ambient material, thus
increasing its metallicity and temperature, which will be reflected in
the X-ray emission.

Studies have shown that in a self-similar case, the R-T fingers also
grow self-similarly. Under adiabatic conditions, whether the expansion
is self-similar or not, the length of the fingers generally does not
exceed more than half the length of the shocked ambient
medium. However, while the growth rate of the instability does not
depend on the shock compression, $\chi$, the width of the shocked region
does. In remnants undergoing efficient particle acceleration the width
of the shock region can decrease to almost half of the case with no
particle acceleration \citep{Decourchelle00}, and therefore the R-T fingers can reach almost
all the way to the FS \citep{Blondin01}, further
complicating the dynamics and kinematics.

\subsection{Shock-Cloud Interactions}

One other generic ingredient in the SNR interaction
is the presence of density enhancements,
particularly within the CSM.  Early on, the role of these
dense ``clouds'' in the CSM was posited to help explain the
observation of {\it both} optical emission due to 
low temperatures (slow, radiative shocks)
as well as X-ray emission due to higher velocities and
temperatures \citep{McKee75,Bychkov75,Stone92,Chugai94}.
In particular this was invoked for the case of the young oxygen-rich SNRs
which showed bright [O III] optical emission along with strong X-ray
emission \citep{Sutherland95b}.

One helpful aspect of the shock-cloud interactions is that the fraction
of material needed to be in clouds is often small and need not greatly disturb
the overall hydrodynamics.  It is therefore possible to focus on just the
shock-cloud interation as a sub-problem of its own \citep{Sgro75,Klein94}.
Using $\chi_c$ to represent the density ratio of the unshocked cloud to the ambient
density, the ``cloud crushing'' time scale is given by the time for the
transmitted shock to travel a cloud radius $a_0$:
\begin{equation}
\tau_{\rm cc} \,\approx\, {{a_0}\over{v_s\,\chi_c^{-1/2}}} ~.
\end{equation}
\noindent where $v_s$ is the shock speed in the ambient (non-cloud) medium.
Of particular interest here is the kinematics of the cloud throughout the
interaction and the resulting state(s) of the cloud material after
the interaction, ending in some combination of evaporated and condensed
(radiative) components.
Thermal conduction can affect the process and enhance the
likelyhood of some or all of the cloud evaporating \citep{Orlando05}.
Studies of the effect of both the geometry and the abruptness of the
cloud boundaries are being made, e.g.,  \citet{Miceli06b} look at
the case of an elliptical cloud.  Looking towards supporting
more detailed observations, a diagnostic that can be applied to the
imaged X-ray emisssion from the remains of a shocked cloud
is presented in \citet{Orlando10}.

What about real shocked clouds?  ~\citet{Klein03} show images of a
laboratory shock-cloud interaction which does resemble the hydro simulations;
these results are invoked to explain a feature seen in a {\it Chandra} Puppis A 
observation \citep{Hwang05} at much larger scales.
Another example is the modeling of part of the Cygnus Loop
with a combination of a smoothly-varying enhanced-density
region which contains discrete higher-density internal clumps \citep{Patnaude05}.

\section{Observations}
\label{sec:obs}

There is a large variety seen in the X-ray images of SNRs,
certainly more than is predicted by simple 1D self-similar models.
This variety is so great that even the fundamental
difference between core-collapse and Type Ia SNe
is often blurred at the SNR stage; though we are making progress
in assessing this key characteristic
from X-ray SNR data directly \citep{Lopez09}.

Following some general comments on X-ray kinematic measurement,
the sections below summarize the \emph{direct kinematic} measurements
that have been made to-date on supernova remnants, both Type~Ia,
Table~\ref{tab:Ia}, and core collapse, Table~\ref{tab:CC}.
Because of this measurement criteria,
only a fraction of the $\sim$\,60 Galactic and $\sim$\,30 Magellanic
cloud SNRs observed by \Chandra\,\footnote{See the Chandra SNR Catalog at
{\tt http://hea-www.cfa.harvard.edu/ChandraSNR/}.}
are mentioned here.  Please consult the cited
works and their references for context, previous work, and
interpretation of the results.

\subsection{Implicit and Direct X-ray Kinematics}

The emission of X-rays from SNRs is in itself an indication of
the underlying kinematics.  In the case of thermal emission
the spectral fitting parameters, $kT_e$ and $\tau$, are
closely related to the hydrodynamics and one can effectively
constrain the kinematics under model assumptions even without
explicitly measuring the velocity of the X-ray plasma.  As a simple example
\citet{Gonzalez03} apply a Sedov model to G292's X-ray emission.
In a more complex analysis, \citet{Laming03} present measurements of fitted
$kT_e$ and $\tau$ values for a coherent set of knots in Cas~A and match
the $kT_e$-$\tau$ pattern with model predictions to constrain kinematics.
There are many SNR X-ray studies of this ``implicit kinematics'' type
and a list of them, never mind their review, is well beyond this author.

The kinematics of SNRs can be directly measured in the X-ray through Doppler velocity
effects in their spectra, via $v_{\rm bulk}$ and/or $\sigma_{\rm therm}$,
and through proper motions seen in imaging
observations taken at multiple epochs, generally giving $v_{\rm FS}\approx v_s$.
These measurements provide
complementary information to the parameters of spectral fits.

\paragraph{\bf Doppler Measurements}
Doppler, or ``radial'', velocity measurements have been common in the
optical for well over a century.  In the X-ray band the first 
measurement of SNR velocities was made spectroscopically using the
\Einstein\ Focal-Plane Crystal Spectrometer (FPCS) observing Cas~A \citep{Markert83}.
Currently the high spectral resolution of the gratings on \XMM\ and \Chandra\
as well as the improved stability of CCD instruments, e.g., on {\it Suzaku},
are providing Doppler measurements on many more SNRs.
The Doppler measurement requires a narrow spectral feature (a line or edge)
and hence is most applicable to sources with thermal emission lines.
Measuring Doppler velocities and line profiles
to an accuracy of 100\,\kms\ or less requires 
some of the highest spectral resolving powers, $E/dE$, shown
on Paerels' plot \citep{PaerelsThisVol}.

\paragraph{\bf Caveat: Imposter Velocities via Ionization State}
Although Doppler measurements can be very ``clean'', it is possible
to get erroneous results when the lines being measured are not
fully resolved.  For example, in Fe-K measurements of Cas~A
\citep{DeLaney10} and N103B \citep{vanderHeyden02},
it is likely that the Fe-K line complex centroid varies not only
with velocity but with temperature and ionization age as well.

\paragraph{\bf Proper Motion Measurements}
Because proper motion measurements require both spatial resolution and a
long time baseline, these measurements came later and the first ones
included data from \Einstein\ and \ROSAT\ \citep{Hughes99,Hughes00b}.
In recent years, \Chandra's sub-arcsecond resolution coupled to its 10$+$-year
lifetime is providing X-ray proper motion measurements of many SNRs.

A useful way of putting a proper motion measurement into a dynamic
context is with the {\it expansion parameter}:
the power-law index $m$ in the expression $r\propto t^m$,
where $t$ is the time since the SN explosion.
This has the useful interpretation of converting
fractional changes in the SNR age to fractional changes in the SNR radius:
$\Delta r/r = m~\Delta t/t$.  For example, for free expansion we have $m=1$
and for an interacting, decelerating dynamics: $m<1$.
Another way to view it, especially for non-monotonic density profiles, is
as the ratio of the current velocity to the historical average velocity:
$v(t) = dr/dt = m~r(t)/t$.

\paragraph{\bf Caveat: Feature evolution}
One complexity with proper motion is that unlike the case of ballistic
optical knots, say, X-ray features are likely to be evolving as well as moving
in time and this needs to be considered in proper motion measurements.
The measured proper motion is the ``motion'' of the brightness distribution
between two epochs - if the feature has evolved in time then the
velocity measured may not be the actual transverse motion of the structure or
material.  This may be more important for measuring shocked ejecta
or CSM clumps, say, as opposed to the narrow non-thermal FS
structures.

\subsection{``Type~Ia'' X-Ray Observations}
\label{sec:Ia}

Type~Ia SNRs that have direct X-ray kinematic measurements are
listed in Table~\ref{tab:Ia} and discussed briefly in turn in the following
paragraphs.
Note that Type~Ia SNe {\it per se} have not been detected in X-rays 
with an early-time ($\sim$\,10 d) upper limit of 
$\sim 2\times 10^{37}$~\ergs\ \citep{Hughes07}.

\paragraph{\bf G1.9+0.3}
This is the youngest known Galactic SNR, and it is still growing in both X-ray size 
as well as in flux.  The diameter increased by 16\,\%\,$\pm$3\% in 22 years between radio (1985) 
and X-ray (2007) measurements,
as confirmed by recent radio and X-ray re-observations \citep{Reynolds08a,Borkowski10}.
Using an SNR diameter expansion from 84 to 97.5
arc seconds gives the FS velocity shown in
Table~\ref{tab:Ia}; detailed X-ray proper motion analyses between the 2007
and 2009 X-ray epochs are ongoing.
Fitting the spectrum of the ``north rim'', \citet{Borkowski10} used a flat-top
(thin shell) model giving $v_{\rm shell}\sim 14\,000$\,\kms, which is expressed
as the equivalent Gaussian value in the table.


\begin{table}
\caption{Type~Ia SNRs: Summary of X-ray Kinematics.  The value $\sigma_v$
represents observed Doppler broadening (combined thermal \& bulk), whereas $v_{\rm FS}$ expresses
the measured proper motion as an equivalent FS velocity.
\label{tab:Ia} }
\begin{tabular}{llcccr}
\hline\noalign{\smallskip}
SNR Name & $\sigma_{\rm v}$\,(\kms) & $v_{\rm FS}$\,(\kms) & Dia.('') & Dist.\,(kpc) & Age (yr)\\
\noalign{\smallskip}\hline\noalign{\smallskip}
G1.9+0.3         & 8100\,$\pm$2900 (N rim) &  12\,400   &  98        & 8.5      & $\sim$\,100 \\
Kepler           &    ---                  & 2200; 3900 & 210  (N-S) & 4\,$\pm$1 & 406 \\
N103B            &      $<$\,350 ($E<1$~keV)  &  ---       & 30         & 50 (LMC) &  1000 -- 2000  \\
SN~1006          & 1775\,$\pm$250 (NW knot)& 5000 (NE)  & 1800       & $\sim$\,2.2 &  1004 \\
SNR~0509-67.5    & 4900\,$\pm$420 (Fe-L, O)  & 6700       & 32         & 50 (LMC)   & $\sim$\,400 \\
SNR~0519-69.0    & 1873\,$\pm$50           & ---        & 33         & 50 (LMC) & 600\,$\pm$200 \\ 
Tycho            & $\sim$\,2600 (Fe-K)     & 2200--4400 & 500        & 2.3\,$\pm$0.8  &    438 \\
\noalign{\smallskip}\hline\noalign{\smallskip}
SN~2002bo &  \multicolumn{2}{c}{Upper limit: 1.8e37 \ergs, 0.5-2~keV} & 0  &
22~Mpc &  at 9.3~d \\
\noalign{\smallskip}\hline
\end{tabular}
\end{table}


\paragraph{{\bf Kepler} (SN~1604, G4.5+6.8)}
\citet{Hughes99} used \Einstein\ and \ROSAT\ data to measure an average expansion rate
of the Kepler X-ray images of 0.239\,\%\,\pyr, giving an expansion
parameter value of $m\approx 0.93$ with an error of order $\pm$0.10.  Note
that the result will be weighted toward the brighter regions, additionally,
this global image-fitting measurement is sensitive to the
``feature evolution'' caveat mentioned above.

\Chandra\ observations of Kepler were made in 2000, 2004, and 2006
and allow a variety of X-ray proper motion measurements.  In the case
of narrow FS features, \citet{Katsuda08b} measured the motion of the 
outer rim using small rectangular regions at selected azimuths;
the values vary between 0.08 to 0.25~\%\,\pyr, with smaller values to
the north.  The shock fronts are roughly circular with radii of 93 (N) and
120 (S) arcseconds and expansion parameters of order 0.5 (N) and 0.7 (S).
These values give the FS velocities shown in the table using the given
distance.

\citet{Vink08} used the \Chandra\ data to make whole-remnant expansion
measurements in distinct energy ranges and six azimuthal sectors.  The N-S 
difference in expansion parameter is also seen, giving values of 0.35 (N) and
0.6 (S). \citet{Vink08} also measures the expansion rate for a narrow
filament in the east, getting a value of 0.176\,$\pm$0.007~\%\,\pyr;
this is in good agreement with the coresponding measurements of
\citet{Katsuda08b}
which give 0.17\,$\pm$0.015 (Reg-4, Reg-5). Hence the difference in expansion
parameters is likely due to the difference in the component that is
being measured: the FS versus the ejecta.

\paragraph{\bf N103B}
Given its small size and presumed large age \citep{Lewis03}, proper motion measurements
in the X-ray will require time baselines of many decades to be significant.
Observed with both \XMM\ and \Chandra\ gratings, N103B has so far
produced only upper limits to a Doppler velocity blur:
the RGS spectrum did not reqire any additional broadening
beyond the spatial contribution \citep{vanderHeyden02}, suggesting
a limit of order $\sigma_v <$ 350~\kms\ for the lines below 1~keV.

\paragraph{{\bf SN~1006} (G327.6+14.6)}
\XMM\ observed the bright metal-rich knot on the northwest rim
of SN~1006, and the RGS data show a dramatic thermal Doppler
broadening of the O\,VII triplet with FWHM $\approx$ 4200\,\kms\
\citep{Vink03}.
Ninety degrees away, the largely non-thermal NE limb shows
a measured proper motion of the FS
of $\sim$\,0.48\,\arcsec\,\pyr\ \citep{Katsuda09}.

\paragraph{\bf SNR~0509-67.5}
\citet{Vink06a} shows a global
velocity broadening of $\sigma_{v}\approx 6500$\,\kms\ in
the O VIII Ly-$\alpha$ line.  Using the bright Fe and O lines in the RGS
spectrum \citet{Kosenko08} measure $\sigma_v\approx$\, 4900\,$\pm$420\,\kms.
\citet{Helder10} mention an initial X-ray proper motion measurement
based on \Chandra\ data which results in $v_s=6700\,\pm400$\,\kms.

\paragraph{\bf SNR~0519-69.0}
\citet{Vink06a} shows a global velocity broadening of $\sigma_v\approx 1700$\,\kms\ in
the O VIII Ly-$\alpha$ line.  Recent analysis by \citet{Kosenko10} gives a
Doppler broadening of $\sigma_v\approx$\, 1873\,$\pm$50\,\kms\ in the RGS spectrum.
They also infer an age of 450 years from the X-ray derived CSM density and
shock radii,
in reasonable agreement with an age of 600 years based on light echoes.

\paragraph{{\bf Tycho} (SN~1572, G120.1+1.4)} \Chandra\ data have provided
beautiful images and detailed spatial-spectral measurements of Tycho \citep{Warren05},
a much studied Type~Ia SNR.
\ASCA\ data showed the emission lines to be broad, possibly as large
as 7500\,\kms\ FWHM \citep{Hwang98}.
Line broadening is seen in Si, S, and Fe-K lines
using the excellent resolution of the {\it Suzaku} XIS CCDs.
As expected for a bulk-plus-thermal combination,
the spatially resolved FWHM shows both a decrease with radius and an excess above
instrumental at the rim \citep{Furuzawa09}.
The center data suggest that the Fe-K-emitting shell is expanding with
an average $v_{\rm bulk}\approx$\,3000\,\kms; the width at the rim corrected for
possible ionization-state effects gives $\sigma_{\rm therm} \approx$
1900\,$\pm$500\,\kms.  These values have been appropriately combined to get the
expected full-SNR $\sigma_v$ given in the table.
The intermediate mass elements (Si, S, Ar) show expansion velocities
$\approx$\,20\%\ greater than Fe \citep{Hayato10}; this is in rough agreement with
a homologous dynamics and the expected/observed
layering of the Type~Ia structure, with Fe toward the interior and
Si, S, and Ar at larger radii.
X-ray proper motion measurements were made using \Einstein\ and \ROSAT\ data \citep{Hughes00c}
and most recently with \Chandra\ data over a 7 year baseline
\citep{Katsuda10a}.  The
values vary with azimuth, in the range 0.2--0.4 arcseconds\,yr$^{-1}$.

\subsection{Core-Collapse X-Ray Observations}
\label{sec:CC}

In constrast to the Type~Ia SNRs, core-collapse SNRs seem to show
greater asymmetry: in their X-ray line emission \citep{Lopez09}, 
in the presence of fast moving outer optical knots of ejecta, 
and at their earliest stages as SNe \citep{WangWheeler08}.
Most of the CC SNRs listed in Table~\ref{tab:CC} are members of the
``oxygen-rich'' class, e.g., showing high velocity [O III] lines
in the optical \citep{Sutherland95b}.  This emission traces the ejecta
dynamics and is therefore relevant to the global X-ray dynamics as well.

\begin{figure*}[t]
\includegraphics[width=0.47\textwidth]{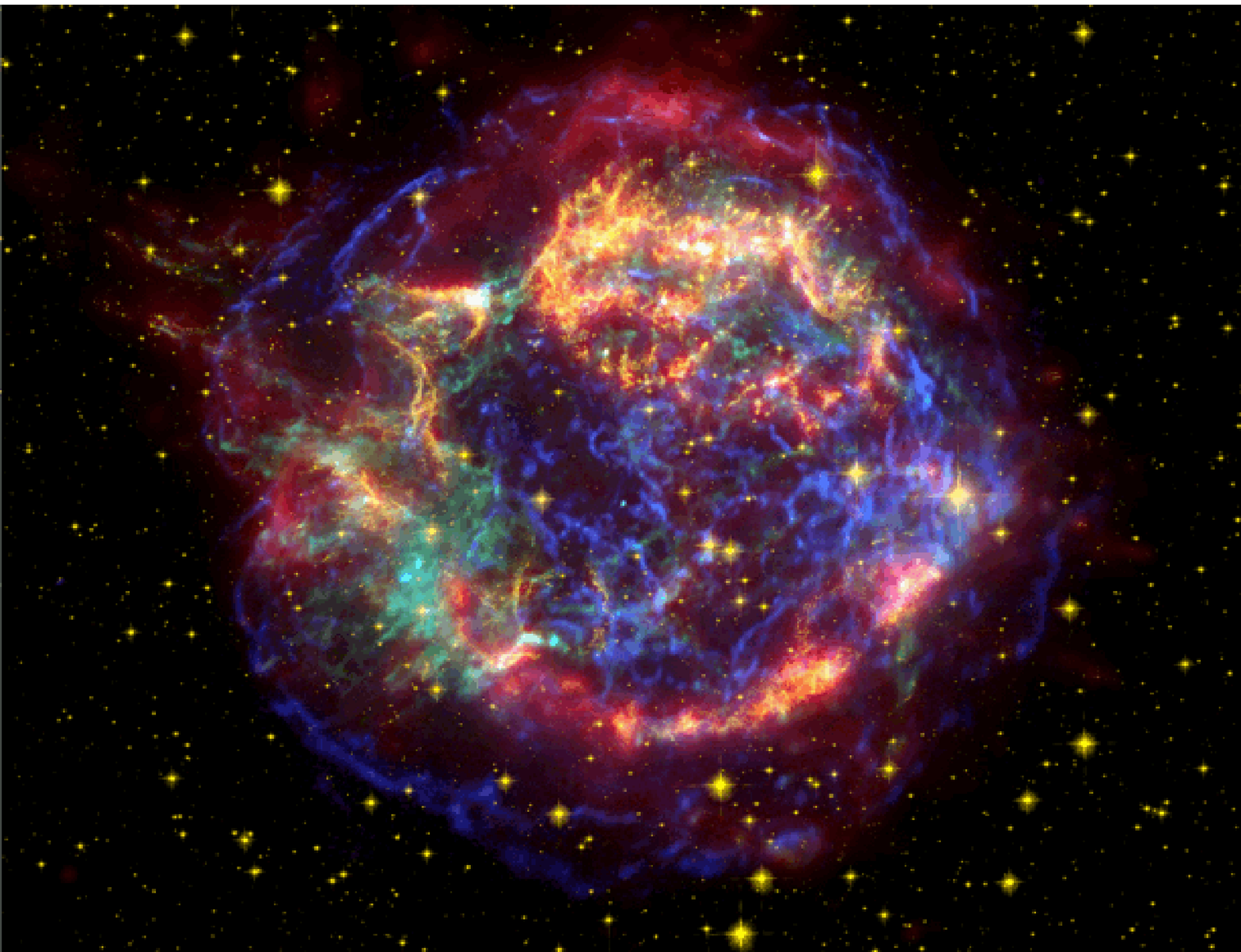}
\includegraphics[width=0.53\textwidth]{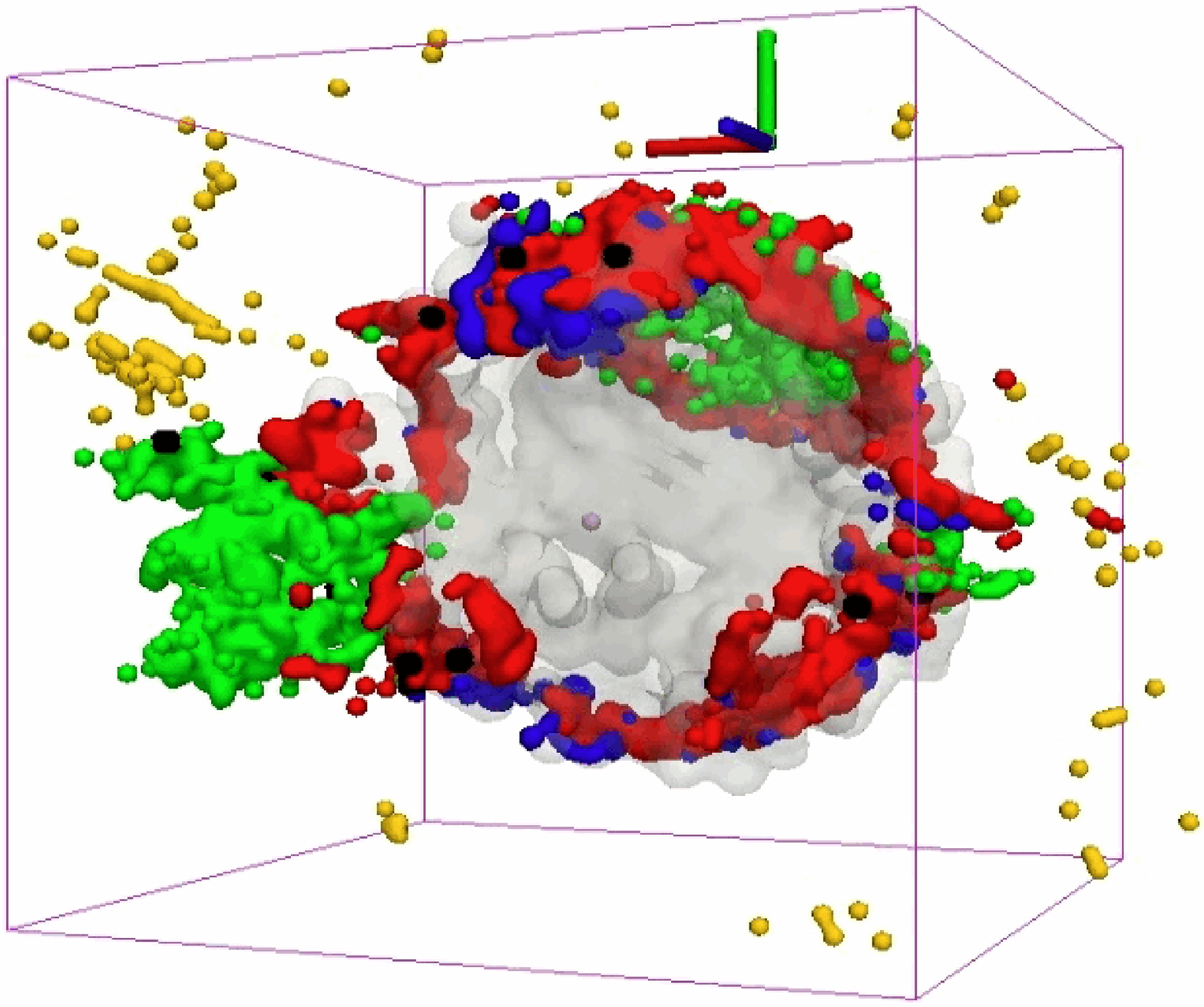}
\caption{Cas~A present. Left: Composite view of Cas~A in X-ray (\Chandra, green \& blue), visible
(HST, yellow), and
the IR (Spitzer, red); from \Chandra\ Photo album, released June 2005.
Right: A detailed 3D reconstruction of Cas~A also in X-ray (black \& green), 
optical (yellow), and IR (red, blue \& gray.)  Doppler shifts measured in X-ray and
infrared lines provided the third dimension, from \cite{DeLaney10}.
\label{fig:CasA3D} }
\end{figure*}

\paragraph{{\bf Cassiopeia~A} (G111.7-2.1)}
Cas~A has been extensively studied because it is nearby (3.4~kpc) and young
($\sim$\,330 yr), has complex structure, and is bright across the
E/M spectrum.
The brightest emission at all wavebands in Cas~A is concentrated
onto the 200\,\arcsec\ diameter Bright Ring, Figure~\ref{fig:CasA3D} (left),
where ejecta from the explosion
are radiating after crossing through and being compressed and heated by the
RS \citep{Morse04,Patnaude07}.  In a few locations, the position of the
RS itself has been identified just inside of the Bright Ring from
the rapid turn-on of optical ejecta \citep{Morse04}.


\begin{figure*}[b] 
\includegraphics[width=0.55\textwidth]{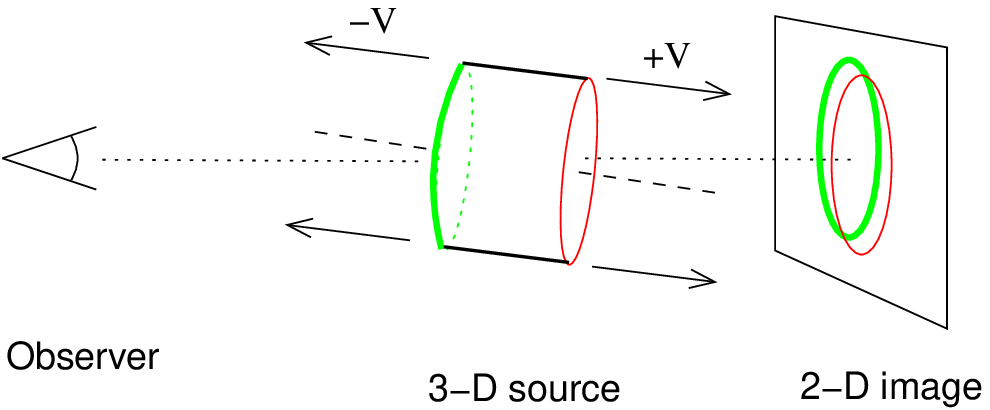}
\includegraphics[width=0.215\textwidth]{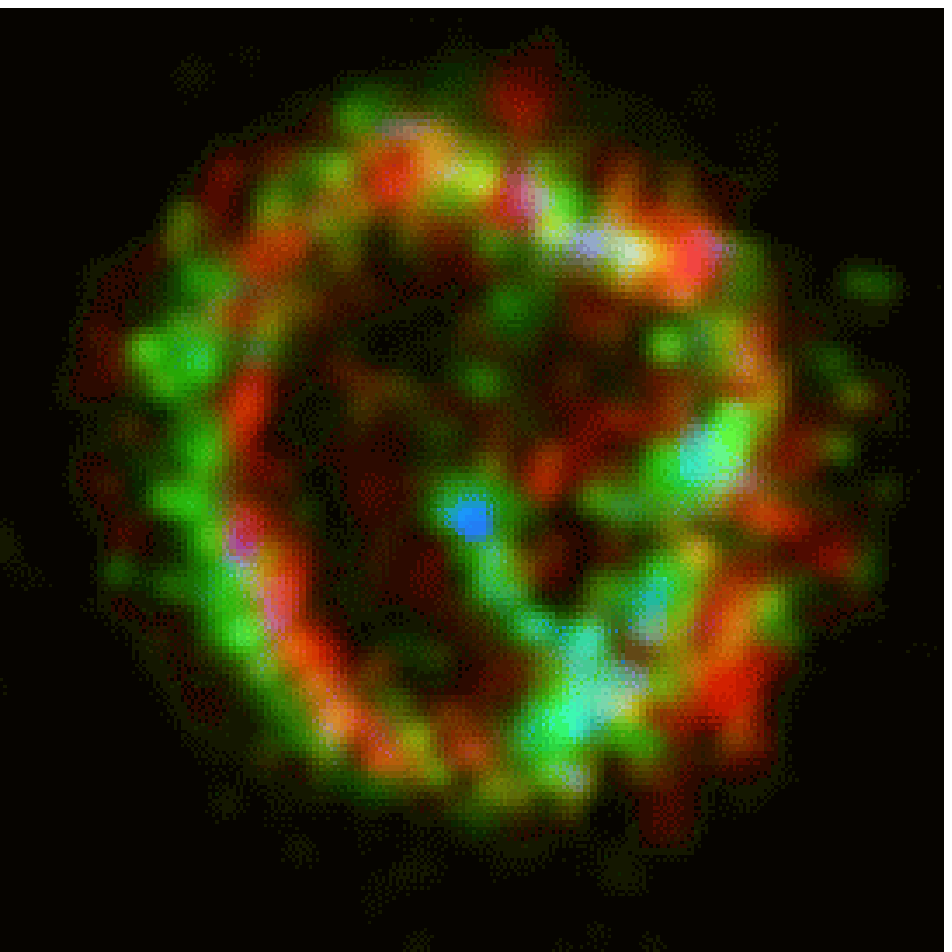}
\includegraphics[width=0.213\textwidth]{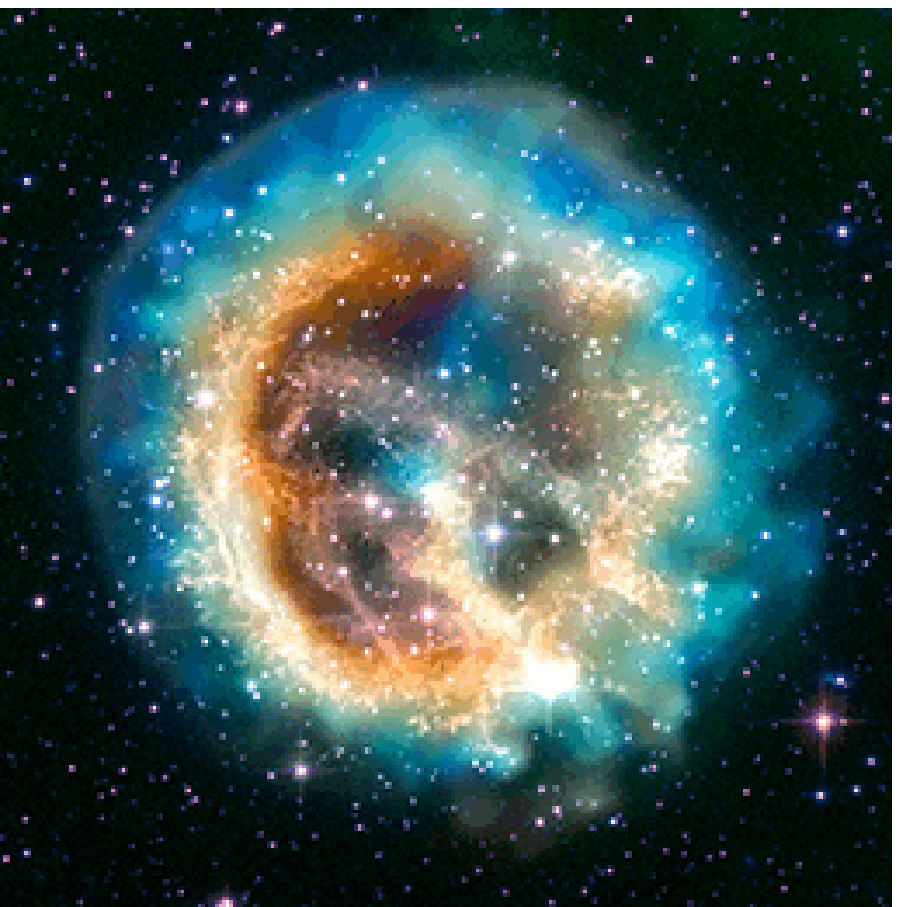}
\caption{The SNR E0102 in the SMC.  At far right E0102 is seen in
false-colored X-rays and optical; from \Chandra\ Photo album, released July 2009.
The middle image of Ne X line emission was created from HETG data and
is color-coded by velocity: +900\,\kms\ (red), $-$900\,\kms\ (green),
$-$1800\,\kms\ (blue); from \citet{Flanagan04}.
The left half shows a cartoon of an underlying 3D structure that can
qualitatively give rise to the spatial-velocity image \citep{Dewey02}.
\label{fig:E0102} }
\end{figure*}


The FS of Cas~A is well defined in \Chandra\ images by a thin
ring of tangential X-ray filaments
at a radius of $\sim$\,153\,\arcsec\ \citep[$t\approx$319\,yr,][]{Gotthelf01,DeLaney03,Hwang04}.
There are also many of these nonthermal filaments projected
across the center of Cas~A (blue in Figure~\ref{fig:CasA3D}, left).
The nature and location of these filaments is not yet clear
-- they may simply be projected FS filaments
\citep{DeLaney04,Patnaude09},
or more intriguingly, they have been modeled as occuring at the RS
as well \citep{Helder08}.

While most of the observed ejecta are concentrated on the
Bright Ring, optical ejecta are also identified at and beyond the
FS \citep{Kamper76,Thor01,Fesen01,Fesen06a}.
Located mostly between 120\,\arcsec\ to 300\,\arcsec\ 
from the center of expansion \citep{Hammell08},
the distribution is aspherical showing jets of Si-rich and S-rich ejecta extending
out to the largest distances in the northeast and southwest.
The optical knots are rapidly expanding, $\sim$\,0.3\,\%\,\pyr\ 
(an expansion parameter $m\approx 1$),
with proper motions of 1000s of \kms\ on
the Bright Ring and up to 14,000\,\kms\ for the optical ejecta at the tip of
the jets \citep{Fesen06b}.  These measurements of the optical ejecta
give an explosion date of A.D.\ 1681$\pm$19.

Spectral variations in the X-ray-emitting ejecta have been mapped by many past
and present X-ray missions \citep{OhashiThisVol}.
The first mappings with the \Einstein\ X-ray
Observatory \citep{Markert83} and \ASCA\ \citep{Holt94} showed large-scale
asymmetries in Doppler structure and lead to a simple inclined-ring model of
the kinematics and 3D structure.  These asymmetries were confirmed with
\XMM\ \citep{Willingale02} and with the
\Chandra\ X-ray Observatory \citep{Hwang01}. \citet{Willingale02}
found that the Si-rich ejecta forms the same general set of ring structures
as the optical emission.  Analyses using
\Chandra\ High Energy Transmission Grating (HETG) data
measure the velocities of individual Si-rich knots, 
with sizes of a few arcseconds, showing velocities 
$v_{\rm Si}$ in the range of $-$3000 to $+$4000~\kms\ \citep{Lazendic06}.
These knot measurements confirmed that the X-ray Doppler expansion rate agrees
with the X-ray proper motion rate, 
$\sim$\,0.2\,\%\,\pyr\ on average \citep{Koralesky98,Vink98,DeLaney04}.
Perhaps somewhat surprisingly, the expansion rate of the FS
is about the same as the X-ray ejecta \citep{DeLaney03,Patnaude09}.

The most recent work by \citet{DeLaney10} combines \Chandra\ (ACIS Fe-K
data and HETG Si data) with Spitzer long-slit spectral imaging (in forbidden
lines of Ne, Ar, and Si).  These components are seen in 3D in Figure~\ref{fig:CasA3D} and
include a garnish of outer optical knots as well.  Spitzer has also been
used to map the center IR emission at high spectral resolution
\citep{Isensee10}.  The gray material in Figure~\ref{fig:CasA3D} identifies Si ejecta
that have not yet reached the RS and thus are unshocked.  When
seen in 3D, the gray material is located on two sheets -- one front and one
back -- and the entire ejecta distribution is flattened, however the RS
is roughly spherical. The Bright Ring of Cas~A is simply the
intersection point between the flattened ejecta and the spherical RS.
The Fe-rich emission is concentrated in three main
areas to the north, southeast, and west. \citet{Hughes00a} suggest an
overturning in the explosion in the southeast, with the Fe emission
extending out to the FS.
However, there are no unshocked Si ejecta immediately interior to the Fe-rich
regions; this may indicate that no overturning of the ejecta has occured.
Instead, the whole ejecta column along that radius may have been displaced
outward with the layer ordering preserved \citep{DeLaney10}.

The small-scale structures in Cas~A evolve on time scales of a few years to
a few decades with ejecta knots and nonthermal filaments appearing,
increasing in brightness, and then fading from view \citep{Patnaude09,
vandenBergh85}.  Additionally, the X-ray ejecta are observed in Cas~A with
a range of ionization states and densities \citep{Lazendic06}, these values
have implications for the further plasma evolution of the features
and we may be able to directly measure the plasma-state changes in future observations.

\paragraph{{\bf E0102} (1E 0102.2-7219)} \Chandra\ imaging
of E0102 (Figure~\ref{fig:E0102}, right-most)
clearly shows the outer blast wave and bright inner rings of H-
and He-like O and Ne emission \citep{Gaetz00}.
A proper motion measurement in the X-ray was made
using \ROSAT, \Einstein, and \Chandra\ data \citep{Hughes00b}, giving 
0.10\,\%\ $\pm$0.025\%~yr$^{-1}$ and the FS velocity in Table~\ref{tab:CC}. 
The shock temperature from eq.\,(\ref{eq:kTshock}) is $kT_s\sim 45$~keV,
and, as mentioned in \S\ref{sec:shock}, we expect $kT_e$ to be below this
value.  \citet{Hughes00b} estimate 2.5~keV as a lower limit based on Coulomb
collisions and the fitted ionization age; this is
greater than the $kT_e\sim 1$~keV seen in the blastwave region
and suggests that some energy must be going into cosmic rays at the shock
\citep{Decourchelle00}.
Recent E0102 optical proper motion measurements \citep{Finkel06}
suggest an age of order 2000 years.

Clear effects of line-of-sight velocities on the order of $\pm$\,1000\,\kms\ are
seen in grating observations of E0102 \citep{Canizares01,Rasmussen01,Flanagan04}.  Spatial-spectral
fitting of the Ne~X line images suggests a cylinder geometry for the X-ray
emission, Figure~\ref{fig:E0102}.   Optical Fabry-Perot and long-slit
observations of the bright [O III] $\lambda$5007\,\AA\ line have shown
similarly-high velocities and suggest a more complex 3D geometry, e.g., see the
references and results presented in \citet{Vogt10}.
E0102's spatial-velocity structure is certainly non-spherical, likely
axisymmetric, and worthy of further study.


\begin{table}
\caption{Core-Collapse SNRs: Summary of X-ray Kinematics.  Specific X-ray
Doppler measurements are given: $v$ indicates bulk velocity and
$\sigma$ is used for broadening measurements.  The value of $v_{\rm FS}$ expresses
the measured proper motion as an equivalent FS velocity.
\label{tab:CC} }
\begin{tabular}{llcccr}
\hline\noalign{\smallskip}
SNR Name  & X-ray Result(s) (\kms) & $v_{\rm FS}$\,(\kms) & 
             Dia.\,('') & Dist.\,(kpc)  & Age (yr)\\
\noalign{\smallskip}\hline\noalign{\smallskip}
Cas~A &  $v_{\rm Si}$: $-$3000 to $+$4000 &  5100  &
               306  & 3.4 &  $\sim$\,329 \\
E0102 & $v_{\rm Ne\,X}\sim\pm$\,1000 & 6200\,$\pm$1500 & 44 & 60 (SMC) &  $\sim$\,2000\\
G292.0+1.8 & $\sigma_{\rm Ne X}\sim$\,1500;~~$\sigma_{\rm O}<$\,730  & ---  &
               530  &  6  & $\sim$\,3000 \\
N132D &  $v_{\rm knot}<$1000; $\sigma_{\rm O}\sim$600 & ---  &
                93 $\times$ 115 & 50 (LMC) &  $\sim$\,3200 \\
Puppis~A & $v_{\rm knot}\sim$\,$-$1700, $-$3400 &  ---  &
                    3300  &     2.2     &   $\sim$\,4000 \\
RCW~86 &  $\sigma_{\rm Fe-K} \sim$\,2200 & 5900\,$\pm$2000  & 2400  & 2.5   & 1825 \\
SNR~0540-69.3 &  $v_{\rm O VIII}\sim$\,$-$2400 &   ---   &
                 65   &   50 (LMC)  &    760 -- 1660 \\
SNR~4449-1 &  ---  & ---   & $\sim$\,0.05 & 3800 &  50 -- 100 \\
\noalign{\smallskip}\hline\noalign{\smallskip}
SN~1987A     &    (Neutrinos!)     &      (\,Radio\,:\,)        &
                           & 50 (LMC)  &  0 \\
~~$\rightarrow$\,1991.6  &  ---   &  (38\,000)        & 
                       (1.3) &   & 4.5 \\
~~$\rightarrow$\,2003.9  & $\sigma_{\rm lines} \sim$\,2300   &  7500 (4700)   &
                         $2r_0\sim$1.48  &   & 17 \\
~~$\rightarrow$\,2010    & $v_{\rm ring}\sim$\,350$\pm$150~[`07]   &   1600\,$\pm$600       &
                         $2r_0\sim$1.56  &   & 23 \\
\noalign{\smallskip}\hline\noalign{\smallskip}
SN~1996cr & $v_{\rm bulk} \sim$\, 4000 &  ---  & 0.01  & 3700 (CG) &  14 \\
\noalign{\smallskip}\hline\noalign{\smallskip}
\end{tabular}
\end{table}


\paragraph{{\bf G292.0+1.8} (MSH 11-54)} This intriguing SNR includes
a pulsar wind nebula and is likely the remains of a 30--40 solar-mass 
star \citep{Park02,Gonzalez03,Park07}.
The radial variation seen in the temperature and emission measure
near the FS prefer models that explode into
an $r^{-2}$ density profile, e.g., into the previous RSG-phase wind medium \citep{LeeJJ10}.

Although G292 is a very large SNR
it does have a bright equatorial belt which has allowed some
\XMM\ RGS spectroscopy:
line broadening is seen for the Ne\,X line with $\sigma_v\sim$\,1500\,\kms,
however, only an upper limit (of 730\,\kms) was placed on broadening in the O lines \citep{Vink04}.
The working hypothesis is that the O is largely from shocked
moderate-density CSM in the belt, while the Ne is from moving ejecta.
Optical velocity measurements in [O III] show fast moving knots within
a shell with $v_{\rm ejecta}\sim 1700$\,\kms\ and confirm the low
velocity of the bar material \citep{Ghavamian05}.
Proper motion measurements of outer optical filaments provide a good age
estimate and show a bi-polar/conical structure \citep{Plunkett09,Winkler09}
 which is similar to a 3D model proposed for {\bf 3C 58} (G130.7+3.1)
\citep{Fesen08}.

\paragraph{\bf N132D} Another beautifully and deeply imaged SNR
\citep{Borkowski07}, N132D has a roughly circular boundary ($r\sim$\,45 arcseconds)
with a ``blow out'' in the NE quadrant.
Early X-ray observations of N132D suggested the explosion
took place in a low-density cavity of a Wolf-Rayet progenitor \citep{Hughes87}.
Based on optical spectroscopy of the high-velocity optical emission \citep{Lasker80},
$v\sim \pm 2000$\,\kms, \citet{Sutherland95a} proposed a complex 3-D axisymmetric
``structural model'' of the SNR.
Further groundbased and HST observations added more ``trees'' to the N132D
forest, including photoionized precursor emission \citep{Morse95,Morse96},
but no clear 3D model has emerged to directly explain the data.

In terms of X-ray spectroscopy, \Einstein\ FPCS data corrected for
spatial-spectral effects \citep{Hwang93} suggested a global O~VIII width in the
range of 1000--2000~\kms\ FWHM; this is
in reasonable agreement with the velocities seen in the optical.
The RGS spectrum of N132D \citep{Behar01} resolved many lines, especially of
Fe, as well as the O VII triplet.
Given the large size, the HETG suffers from spatial-spectral confusion
\citep{Dewey02}, especially given the many Fe lines; although HETG gives a relatively clean
monochromatic image in the O VIII line \citep{Canizares01}.
Focussing on individual knots with the HETG, none of those measured
show Doppler velocities above 1000\,\kms\ \citep{DeLaney07}, somewhat
surprising given the high optical velocities.
Most recently the COS on HST observed an O-rich region (2.5'' diameter)
of N132D in the far-UV lines of O III, O IV, and O V, showing velocity
components at $\sim$\,200 and $\sim$\,800 \kms\ \citep{France09}.
There is certainly much more to measure and model in N132D.

\paragraph{{\bf Puppis~A} (G260.4-3.4)}
Given its large diameter, $\sim$\,1 degree, it's hard to get a full view of
Puppis~A: the X-ray reference image for Puppis~A is still a \ROSAT\ HRI mosaic
that was presented in the paper that confirmed it harbored an unresolved
central stellar remnant, RX J0822--4300 \citep{Petre96}.
This image is usefully combined with radio data as presented in \citet{Castelletti06}.
Puppis~A  shows a roughly rectangular shape with sharp edges along the NE and NW.
The bright eastern knot (BEK) is a prominent feature, likely the interaction
of the SNR shock with an interstellar cloud, and was
the target of a \Chandra\ observation \citep{Hwang05}.
A recent mosaic of \XMM\ and \Chandra\ observations covers almost all of the
SNR giving a three-color image \citep{Katsuda10b}, and a
Suzaku mosaic dataset was used to construct line images and parameter maps
covering most of the remnant \citep{Hwang08}.

The observation of fast moving oxygen filaments in the optical 
puts Puppis~A in the O-rich category \citep{Winkler85}.  Given its large
size, its bright O, Ne, and Fe XVII lines are still best resolved
by the \Einstein\ FPCS using almost $\sim$\,300\,ks of data \citep{Winkler81},
although the implications of the measured fluxes have evolved over the decades.
Observations with \XMM\ show Doppler shifts for fast moving X-ray knots,
$v_{\rm knot}\sim$\, -1700 and -3400 \kms, at locations near the [O III]
``omega'' filament \citep{Katsuda08a}.
One expects that continued X-ray and optical measurements of this SNR will bring out its
full 3D structure.

\paragraph{{\bf RCW~86} (SN~185, G315.4-2.3, MSH 14-63)}
This large SNR is well imaged in an \XMM\ mosaic \citep{Vink06b}
and shows both thermal and non-thermal
emission: detailed studies have been carried out in the bright southwest
``knee'' region \citep{Borkowski01RCW86,Ueno07} and along the E-NE-N rims
\citep{Vink06b}.
A concise summary of RCW~86 and its observations is given
in \citet{Aharonian09} where its imaging in very high
energy ($E>100$\,GeV) $\gamma$-rays is presented.

In the southwest, analysis of Suzaku data
shows emission in the Fe-K range from low-ionization Fe-rich plasma at $\sim$\,6.4
and $\sim$\,7.1~keV; the 6.4~keV line is clearly broadened with $\sigma\sim$\,47~eV
although the mechanism is not clear and may involve explosion within a cavity \citep{Ueno07}.
In the northeast, a \Chandra\ X-ray proper motion measurement over 3 years
gives a value of 0.5~$\pm$0.17 ''\,yr$^{-1}$ indicating a high shock
velociy inspite of the lower post-shock temperatures measured: the conclusion
is that $\ge$\,50\,\% of the post-shock pressure here is due to
cosmic rays \citep{Helder09,Vink10}.

\paragraph{\bf SNR~0540-69.3}
As in the case of the Crab, the pulsar PSR~B0540-69.3 \citep{Serafimovich04}
and the pulsar wind nebula (PWN) of SNR~0540-69.3
are the eye-catchers in X-ray images, e.g., see the deep Chadra observation
of \citet{Park10}.
A ring of [O III] emission with radius $\sim$\,3~arcseconds had been seen from
ground-based observations and is clear in HST images \citep{Morse06}.
Linewidths in the optical give shell expansion velocities of order 1500\,\kms\ 
\citep{Kirshner89,Morse06}.
An observation using the \XMM\ RGS showed emission lines, including
O~VIII, from the western side of the SNR, with an assigned Doppler shift
of $-$2370\,\kms\ \citep{vanderHeyden01}. No \Chandra\ gratings observations
have been made of `0540 to target these thermal emissions because
of the long exposures required.

\paragraph{{\bf SNR~4449-1} (NGC 4449 SNR)}
Although no direct X-ray kinematic measurements have been made of
this SNR in NGC 4449 it is included here because of its O-rich character,
showing velocities $\sim$\,3500\,\kms.
A \Chandra\ X-ray spectrum of SNR~4449-1 shows lines of O, Ne, Mg and Si
roughly similar to those of G292 \citep{Patnaude03}.
Ground-based and HST optical images and spectra presented
by \citet{Milisav08} suggest that the SNe
is interacting with dense CSM from the progenitor, or perhaps
with wind-loss from nearby stars.  Their line profiles
show discrete ``minor peaks'' around $\pm$\,1600\,\kms; this may suggest 
``a possible ring or jet distribution''.
Combining the maximum optical expansion velocity of 6500\,\kms\ with
a VLBI-measured radius gives an age of 50 -- 100 years \citep{Bietenholz10}; additionally
the asymmetric VLBI image suggests a barrel shape
seen in projection, reminiscent of the E0102 cartoon,
Figure~\ref{fig:E0102}.

\begin{figure*}
\center{\includegraphics[width=0.75\textwidth,height=0.45\textwidth]{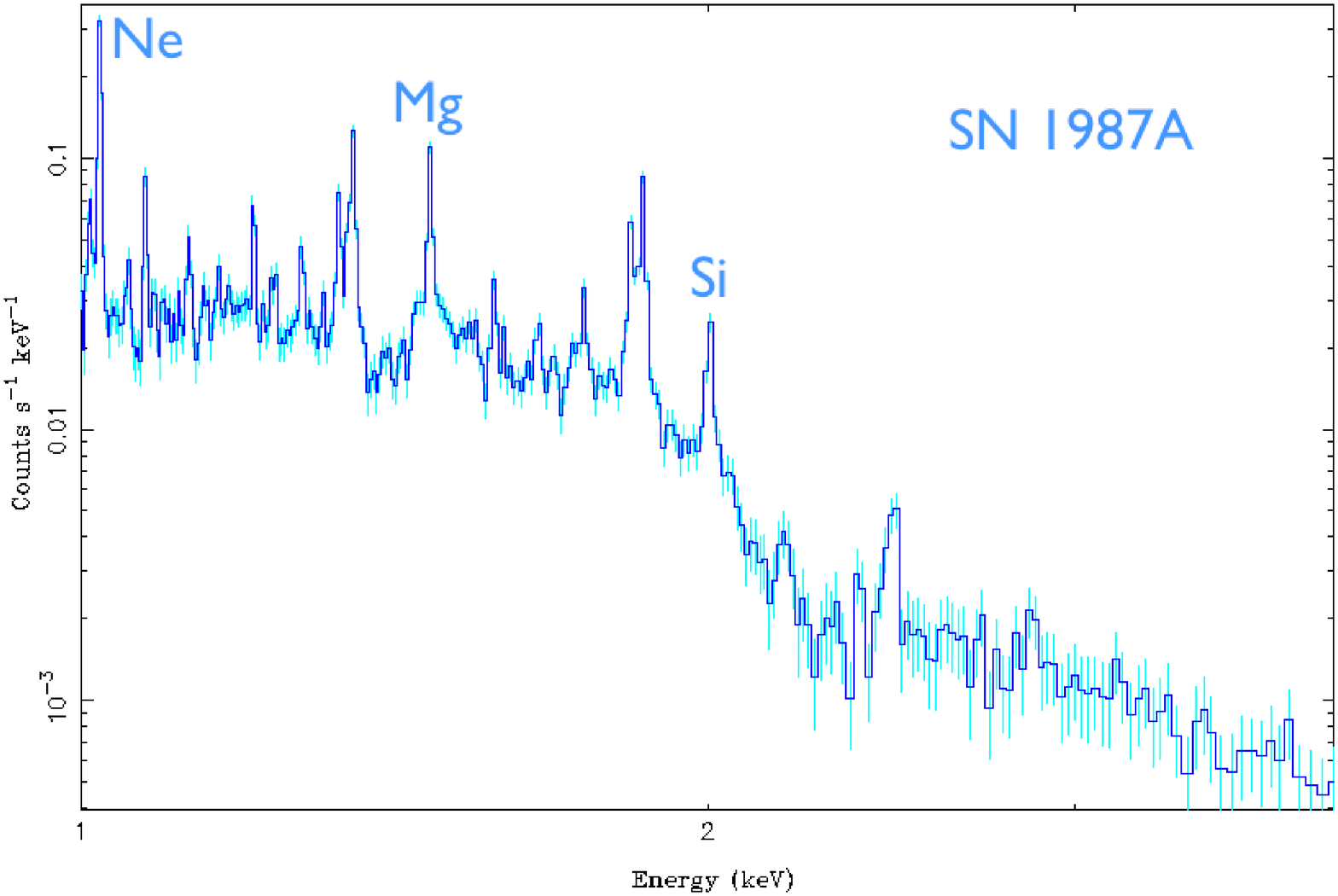}}
\vspace{-0.3cm}
\center{\includegraphics[width=0.75\textwidth,height=0.45\textwidth]{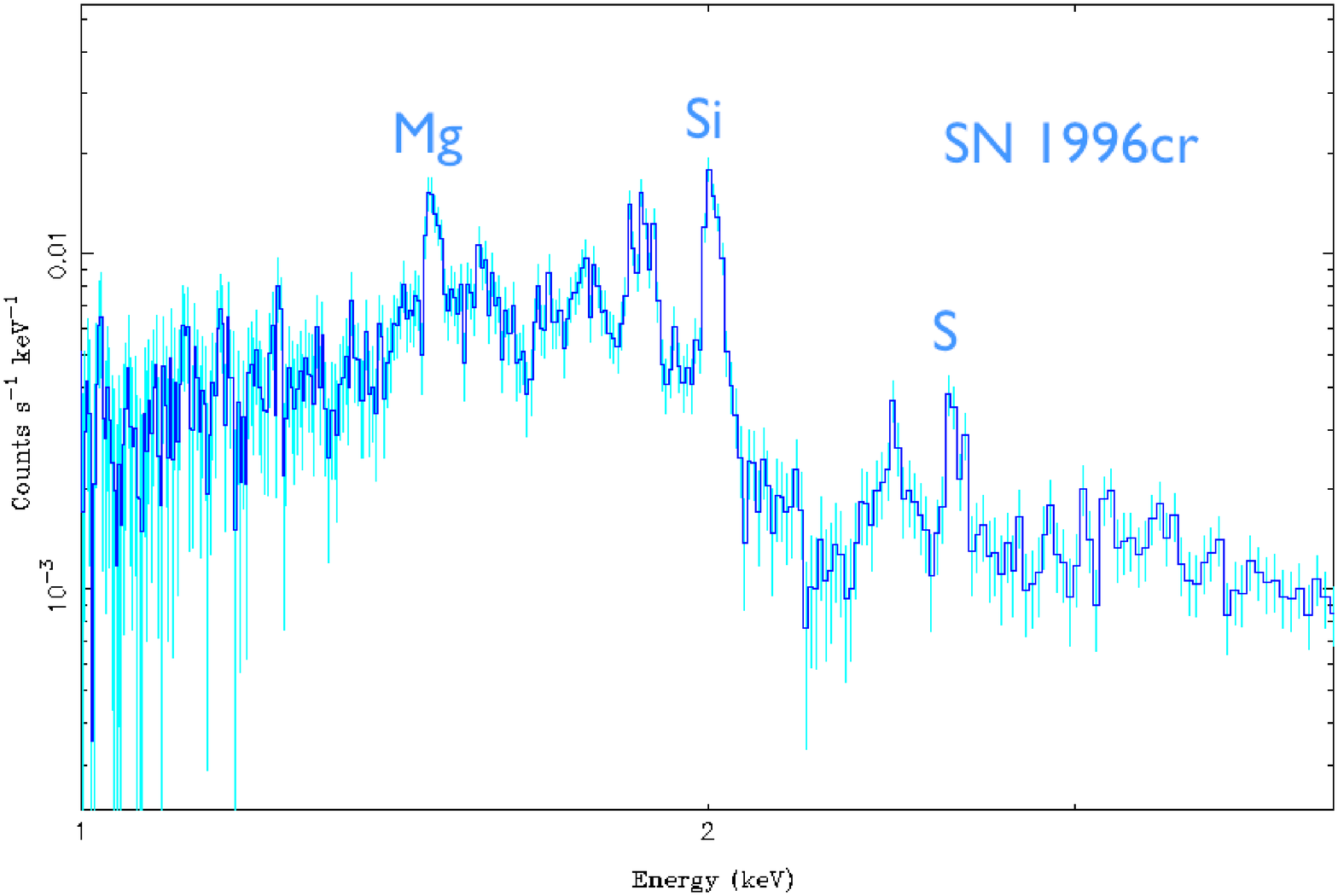}}
\caption{HETG/MEG spectra of early SNe-CSM interations in the 1 to 4 keV range.
{\it Top panel}: SN~1987A (356~ks in 2007) shows many lines from its $\sim$\,1~keV thermal emission;
the lines are somewhat broadened by the 1.5\,\arcsec\ spatial size and
the 100's of \kms\ of Doppler motion.
{\it Bottom panel}: SN~1996cr (489~ks at $\sim$\,2009) 
shows a much higher temperature spectrum ($\sim$\,13~keV) with
clear line broadening due to the $\sim$\,5000\,\kms\ bulk velocities of the shocked material.
{\it Note}: the spectra shown here were obtained with just a few key strokes and mouse clicks
using the {\it TGCat} web archive of
\Chandra\ grating data, {\tt http://tgcat.mit.edu/}~.
\label{fig:SNeHETG} }
\end{figure*}

\subsection{CC SNe observations - The early years}

A large fraction of core-collapse SNe show X-ray
emission within the first 1000 days, with an
$L_X$ that decreases in time \citep{Schlegel95,Immler03}.
This behavior can be modeled as the
interaction of the ejecta with a CSM due to the progenitor's late-phase wind,
$\rho\propto r^{-2}$, often producing a radiative RS and
associated cooling region \citep{Fransson96,Nymark06}.
A well studied example is SN~1993J (in M81, $\sim$4\,Mpc) which so far has data
covering 10 days to 15 years since the explosion \citep{Chandra09}.

A very illuminating look at the transition from SNe to SNRs is seen
in Figure~2 of \citet{Immler05}.  This shows SNe observed at ages
less than 30 years having $L_X$'s generally in the
range $10^{38}$ to $10^{40}$~\ergs\ and decreasing with time.
SNRs with ages 300 to 3000 years are seen to have 
$L_X$ values in the $10^{36}$ to $10^{38}$~\ergs\ range.
The figure makes clear the gap in our observations of core-collapse
SNe/SNR at ages of 30 to 300 years, which we can expect to fill in
with observations, e.g., SNR~4449-1, and patience.

Two SNe outliers in this picture are SN~1987A and SN~1996cr
which each show an {\it increasing} X-ray flux at $\sim$\,5 years
post-explosion.  In these cases
it appears that the CSM consists of a low density region
and then farther out in radius there is dense, structured CSM from the progenitor.
In this section we take a look at SN~1987A and its ``wild cousin'', SN~1996cr,
summarizing their kinematics as seen in high resolution X-ray spectra.

\paragraph{\bf SN~1987A}
For an introduction, see the article by \citet{McCray07} in
the conference proceedings from SN~1987A's 20$^{\rm th}$ birthday. 
The UV flash of the SN caused the CSM around `87A to glow, showing
its triple-ring structure.  Initial expansion of the FS took place in
low density CSM at high velocity \citep{Gaensler07} until $\sim$\,1992.
At this point the FS encountered denser CSM slowing the shock and giving
increased radio and X-ray emission; this phase is dominated
by interaction of the FS with an extended H~II region with $n\approx$100~cm$^{-3}$ 
\citep{Chevalier95}.  The X-ray emission was dominated by
a very broad component seen with the \Chandra\ HETG \citep{Michael02}.
Spectra taken with the \XMM\ RGS in May 2003 are near the end of
this H~II phase \citep{Sturm10}.

In 2004 the FS reached protrusions around the equatorial ring
having $n\approx$10$^4$~cm$^{-3}$\,:
the soft X-ray emission increased more quickly as the X-ray diameter growth slowed.
A full X-ray light curve exists with initial X-ray observations by \ROSAT\ 
and continuing into the current era with a gap of just under 2 years \citep{Haberl06}.

Doppler motion of the shocked ring material was measured through clever observations
with the \Chandra\ LETG \citep{Zhekov05} which indicated lower-than-expected
bulk velocities, confirmed with higher-resolution HETG observations \citep{Dewey08}.
The X-ray image of `87A has been observed at 6 month intervals
with \Chandra\ \citep{Burrows00,Park06} and recent image analyses
give expansion velocities based on model fits to the images \citep{Racusin09},
see Table~\ref{tab:CC}.
Observations of `87A continue to be carried out in all wavebands
as we work to understand and predict the evolution of this wonderful system.

\paragraph{\bf SN~1996cr}
Initially identified as a ULX in the Circinus Galaxy, archival
data showed SN~1996cr was non-detected in X-rays at $\sim$ 1000 days
yet brightened to $L_x \sim 4\times 10^{39}$\,\ergs\ (0.5-8 keV) after 10 years
\citep{Bauer08}.
A 1-D hydro model of the ejecta-CSM interation was created and produces good
agreement with the measured X-ray light curves and spectra across multiple epochs
\citep{Dwarkadas10}.  
SN~1996cr was most likely a massive star, $M > 30$ solar masses,
which went from an RSG to a brief W-R phase before exploding within its $r\sim 0.04$ pc
wind-blown shell.
Doppler line broadening is clearly seen in the
HETG data, Figure~\ref{fig:SNeHETG}, bottom panel, with widths that are in
good agreement with velocities in the hydro-model.  Further analysis of the HETG
observations should allow detailed line-shape fitting of a handful of
bright Si and Fe lines to help constrain the emission geometry.

\begin{figure*}
\includegraphics[width=0.52\textwidth]{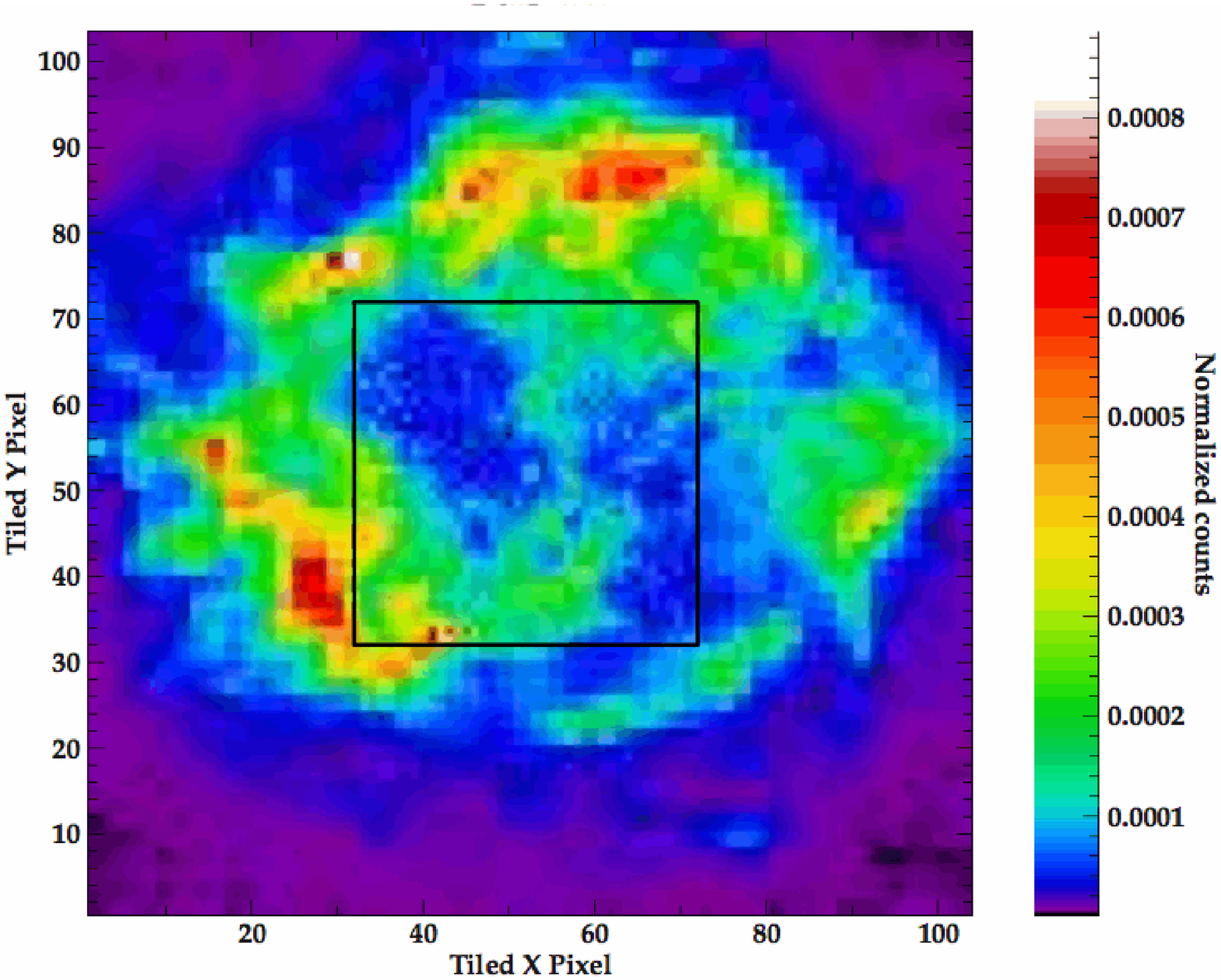}\hspace{0.03\textwidth}
\includegraphics[width=0.44\textwidth]{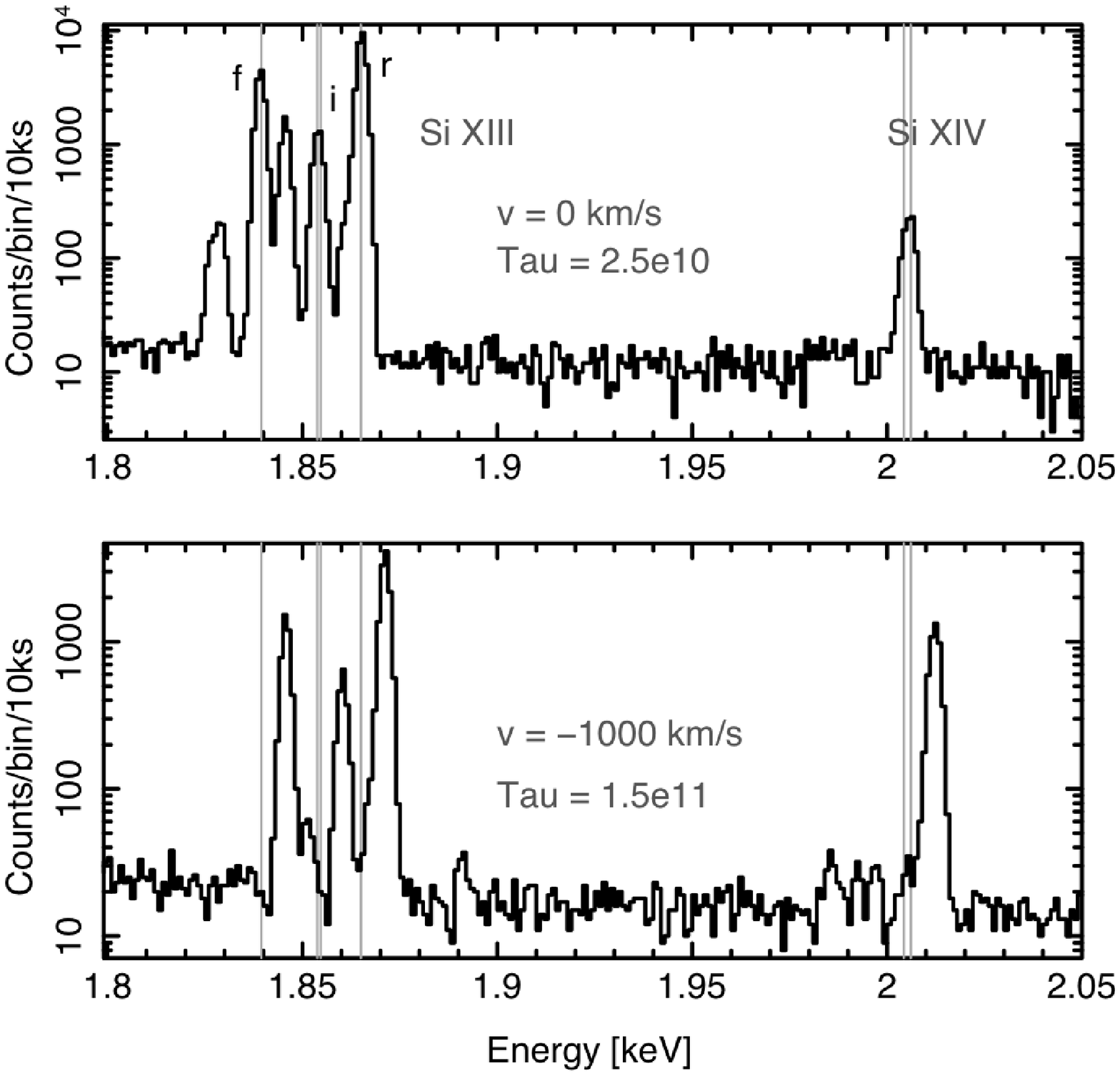}
\caption{Cas~A future observation. {\it Left}: A simulated image of Cas~A using a
microcalorimeter array as envisaged for IXO; from \citet{Davis10}.  {\it
Right}: Simulated spectra from individual regions of the image showing the Si-line region
seen at 2.5~eV resolution  The upper spectrum has $v=0$ and a low ionization
age; the lower spectrum shows 1000\,\kms\ blue-shifted lines from a more highly ionized plasma.
\label{fig:CasAIXO} } 
\end{figure*}

\section{Future X-Ray Obsevatories and SNRs}
\label{sec:future}

A major advance in SNR observations will be made as spectral imaging
with high resolution non-dispersive detectors becomes available,
Figure~\ref{fig:CasAIXO}; for further details
see \citet{OhashiThisVol}.
This new capability will be welcome in the X-ray domain and
will allow studies of Galactic and
Magallanic cloud SNRs with 3D detail exceeding our current astounding view
of Cas~A \citep{DeLaney10}.

There is also a continued need for X-ray grating spectrometers in the low-energy
range, 0.3 to 1.0~keV, see \citet{PaerelsThisVol}.  The
CAT gratings~\citep{Heilmann09} represent a promising technology improvement
with the weight and alignment advantages of a transmission grating like the
HETG \citep{Canizares05} along with the higher efficiency of a reflection
grating, e.g. the \XMM\ RGS.  Such a grating instrument
complements the microcalorimeters by measuring velocities and thermal Doppler
broadening for low-Z and Fe-L ions.
Although optimized for a point source (SNe, GRBs),
the spatial extent of a somewhat extended source can be included in the
spectral analysis as has been demonstrated with \XMM\ and \Chandra\ grating
data \citep{Rasmussen01,Dewey02,Vink03}.

Future kinematic measurements in all wavebands,
including GWs and neutrinos,
will be combined with multi-dimensional simulations, modeling, and analyses
\citep[e.g.,][]{Dewey09,Steffen10}
to provide a rich picture of SNe/SNRs in the decades ahead.

\begin{acknowledgements}
I deeply appreciate receiving suggestions and input from 
Vikram V. Dwarkadas, especially regarding \S\ref{sec:theory}.
I thank Tracey DeLaney for input, helpful discussions,
and an advance copy of \citet{DeLaney10}.
Thanks to colleagues Vikram Dwarkadas and Franz Bauer for
ongoing stimulating discussions.  
Support for this work was provided by NASA/USA through the
Smithsonian Astrophysical Observatory (SAO)
contract SV3-73016 to MIT for Support of the \Chandra\ X-Ray Center,
which is operated by SAO for and
on behalf of NASA under contract NAS8-03060. 
\end{acknowledgements}


\end{document}